\documentclass[journal=jacsat,manuscript=article]{achemso}

\usepackage{graphicx}
\usepackage{latexsym}
\usepackage{stmaryrd}
\usepackage{amsmath}
\usepackage{amssymb}
\usepackage{epstopdf}
\usepackage[colorlinks,linkcolor=blue,anchorcolor=blue,citecolor=red]{hyperref}

\usepackage[T1]{fontenc}

\author{Bing-Lan Wu }
\affiliation{School of Physical Science and Technology, Soochow University, Suzhou, 215006, China}
\author{Zi-Bo Wang}
\affiliation{College of Physics and Electronic Engineering, Sichuan Normal University, Sichuan, 215006, China}
\author{Zhi-Qiang Zhang }\email{zhangzhiqiangphy@163.com}
\affiliation{School of Physical Science and Technology, Soochow University, Suzhou, 215006, China}
\author{Hua Jiang}\email{jianghuaphy@suda.edu.cn}
\affiliation{School of Physical Science and Technology, Soochow University, Suzhou, 215006, China}
\affiliation{Institute for Advanced Study, Soochow University, Suzhou 215006, China}
\date{\today}

\title[An \textsf{achemso} demo]
{Building programable integrated circuits through disordered Chern insulators}


\begin{document}
	\begin{abstract}
		We study the construction of programable integrated circuits with the help of disordered Chern insulators (CIs) in this letter. Specifically, the schemes for low dissipation logic devices and connecting wires are proposed.
		We use the external-gate-induced step voltage to construct spatially adjustable channels, where these channels take the place of the conventional wires.
		Our numerical calculation manifests that the external gates can be adopted to program the arbitrary number of wires ($n$-to-$m$ connections).
		We find that their electron transport is dissipationless and robust against gate voltage fluctuation and disorder strength.
		Furthermore, seven basic logic gates distinct from the conventional structures are proposed.
		Our proposal has potential applications in low power integrated circuits and enlightens the building of integrated circuits in topological materials.
	\end{abstract}

\section{Introduction}
\begin{figure}
	\centering
	\includegraphics[width=0.7\textwidth]{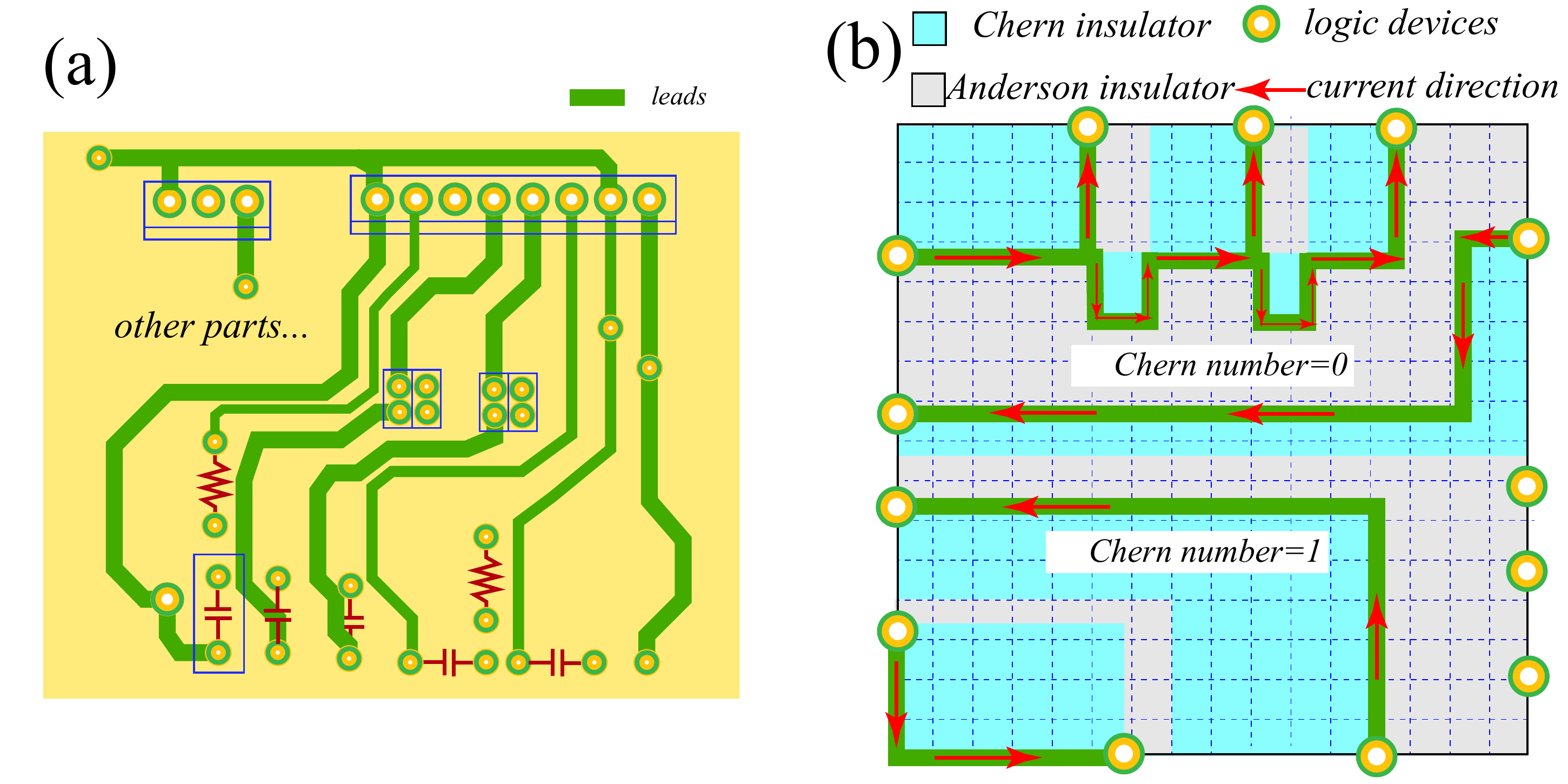}
	\caption{(Color online). (a) Schematic plot of a conventional printed circuit board. The orange areas are the background. The solid green lines are the wires, and the logic devices are marked in yellow circles. (b) Schematic plot of a programmable integrated circuit based on disordered-CIs. The central region is divided into several cyan and gray blocks. The red arrows denote the current directions.}
	\label{f1}
\end{figure}

The integrated circuits made up of metallic wires and logic devices have significantly improved information processing efficiency and dramatically rebuilt our lifestyle \cite{01,02,03}. However, the power dissipation for such devices is one of the most focused challenges, where lots of energy is wasted due to the existence of resistance. In conventional printed circuit boards [see Fig. \ref{f1}(a)], the connections between electronic components are exclusively determined by conductor pattern (metallic wires), in which electron transport is inevitably dissipated \cite{04,05,06}. Furthermore, the complicated structure of logic devices also suffers from heavy joule heat, which also induces overheating troubles \cite{07,08}. In order to overcome those problems, topological insulators have attracted great interest over the past decades \cite{09,10,11,12,13,14,15,16}. These systems are predicted to possess the dissipationless topological edge states and are considered as candidates of ideal wires in integrated circuits \cite{17,18}. Nevertheless, the edge state always sits at the boundary, and its shape is only determined by the geometry of the sample, which limits its applications.

Fortunately, the conducting channels also emerge at the interfaces between two topologically distinct materials \cite{19,20,21}. Especially at the interfaces between quantum anomalous Hall phases with different Chern numbers, the one-dimensional chiral states with dissipationless transport are available \cite{22,23}. Compared to the topological edge states, the interface states exhibit higher tunability, engineered spatially. Recently, we proposed the existence of a chiral interface state with quantized transport in disordered CIs \cite{24}. Generally, the Hall conductivity $ \sigma_{xy} $ for a disordered CI is $ e^{2}/h $ inside the mobility gap, while sharply jumps to $ 0 $ at two mobility edges since all the bulk states are localized by Anderson disorder \cite{24,25,26,27,28,29,30}. When a step potential is adopted by an external gate, the Chern number of two adjacent areas (separated by external gates) can be zero and one, respectively. Hence, the dissipationless chiral channels, which are spatially adjustable by external gates, emerge at the interface \cite{24}.

In this letter, we show that the interface channels with arbitrary trajectories can be constructed by controlling the external gates in disordered CIs, and its robustness against backscattering is also preserved. Utilizing such perfect transport properties, we propose a programable circuit board. As shown in Fig. \ref{f1}(b), the central region of the sample is divided into several blocks, among which appropriate gate voltage arrangements are considered to program the required wires to connect the corresponding devices. Our numerical results manifest that the partition of current can also be realized in this programable circuit, and the number of branch wires can be adjusted by external gate voltage manipulation. Significantly, the programable chiral interface channels can also be adopted to construct all seven basic logic gates\cite{31,32,33,34}. These structure-simplified logic gates are compatible with wires, which take full advantage of disordered CIs' topological nature. Our proposal of dissipationless wires and basic logic gates provide a route to build integrated circuits in topological systems.

\section{Result}
Our investigation is based on Qi-Wu-Zhang CI model \cite{35}, and the Hamiltonian in the square lattice reads:
\begin{align}
	\begin{split}
		H=&\sum_i[c_{i}^{\dagger}(\frac{t\sigma_{z}}{2}-iv\sigma_{y})c_{i+\hat{x}}+c_{i}^{\dagger}(\frac{t\sigma_{z}}{2}-iv\sigma_{x})c_{i+\hat{y}}+h.c.]\\
		&+\sum_i[c_{i}^{\dagger}(m-2t)\sigma_{z}c_{i}+c_{i}^{\dagger}(V_{i}+W_{i})\sigma_{z}c_{i}]
	\end{split}
\end{align}
where $c_{i}^{\dagger}$ ($ c_{i} $) is the creation (annihilation) operator on site $i$. $ \sigma_{x,y,z}$  are Pauli matrices and $ \sigma_{0}$ is the $ 2\times 2 $ identity matrix.
The parameters are fixed at $ v=0.5t $, $ m=t $, where $ v $, $ m $ and $ t $ are Fermi velocity, mass and hopping energy, respectively.
$ V_{i} $ shows the profile of the gate-induced potential, and $ W_{i} $ is the Anderson disorder uniformly distributed within $ [-\frac{W}{2},\frac{W}{2}] $. $ W $ is the disorder strength.

\begin{figure}[t]
	\centering
	\includegraphics[width=0.8\textwidth]{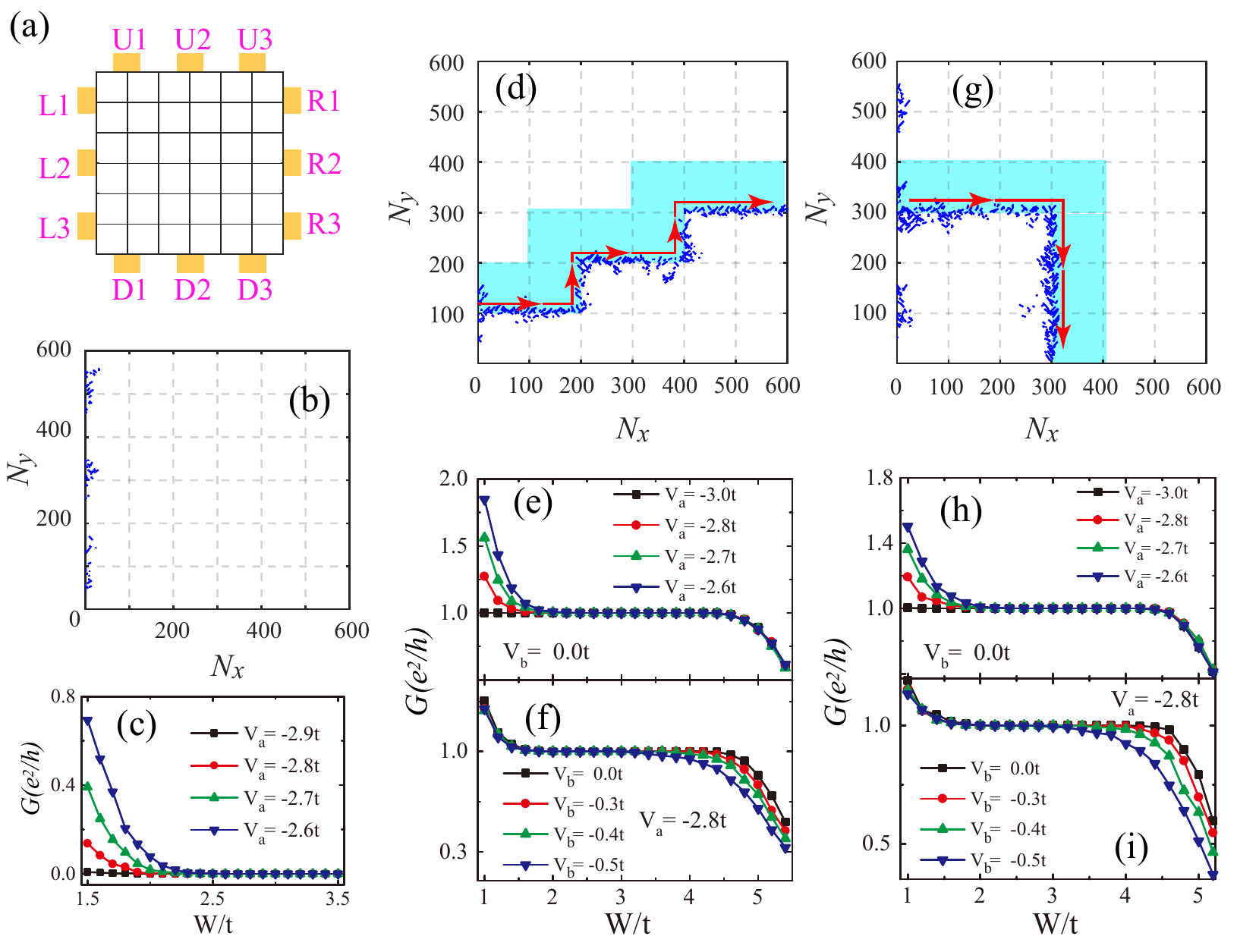}
	\caption{(Color online). (a) Schematic of programable circuit device. The central region is divided into $ 6\times6 $ blocks with size $ 100a\times100a $, where $ a $ is lattice constant. (b)(d)(g) Local current density distribution corresponding to (a) with disorder strength $ W=3.5t $ under different gate voltage configuration. Here, the gate voltage on white and blue blocks are $ V_{a} $ and $ V_{b} $, which take the standard values $ V_{a}=-2.8t $,~$ V_{b}=0 $ if not specified. The red arrows denote the chiral interface channels. (c) Differential conductances $ G $ of case (b) versus $ W $ under different $ V_{a} $. (e)-(f) and (h)-(i) are similar to those in (c), except $ G $ for cases (d) and (g), respectively.}
	\label{f2}
\end{figure}

The wires and the logic gates are basic building blocks for conventional integrated circuits. We first propose the realization of programable `wires' with the help of disordered CIs. The transport properties of a typical device are studied as illustrated in Fig. \ref{f2}(a). The central region is divided into $ 6\times6 $ little blocks, where each block is attached with an external gate. The blocks' potential can be manipulated by the corresponding gates independently. To simulate the typical cases in printed circuits boards, the gate voltage takes two discrete standard values $ V_{a} $ and $ V_{b} $ for white and blue blocks, respectively. Furthermore, 12 logic devices are considered, which are labeled as ($ U_{1},U_{2},U_{3} $), ($ D_{1},D_{2},D_{3} $), ($ L_{1},L_{2},L_{3} $), and ($ R_{1},R_{2},R_{3} $) connecting up, down, left and right boundary, respectively.

\begin{figure}[t]
	\centering
	\includegraphics[width=0.8\textwidth]{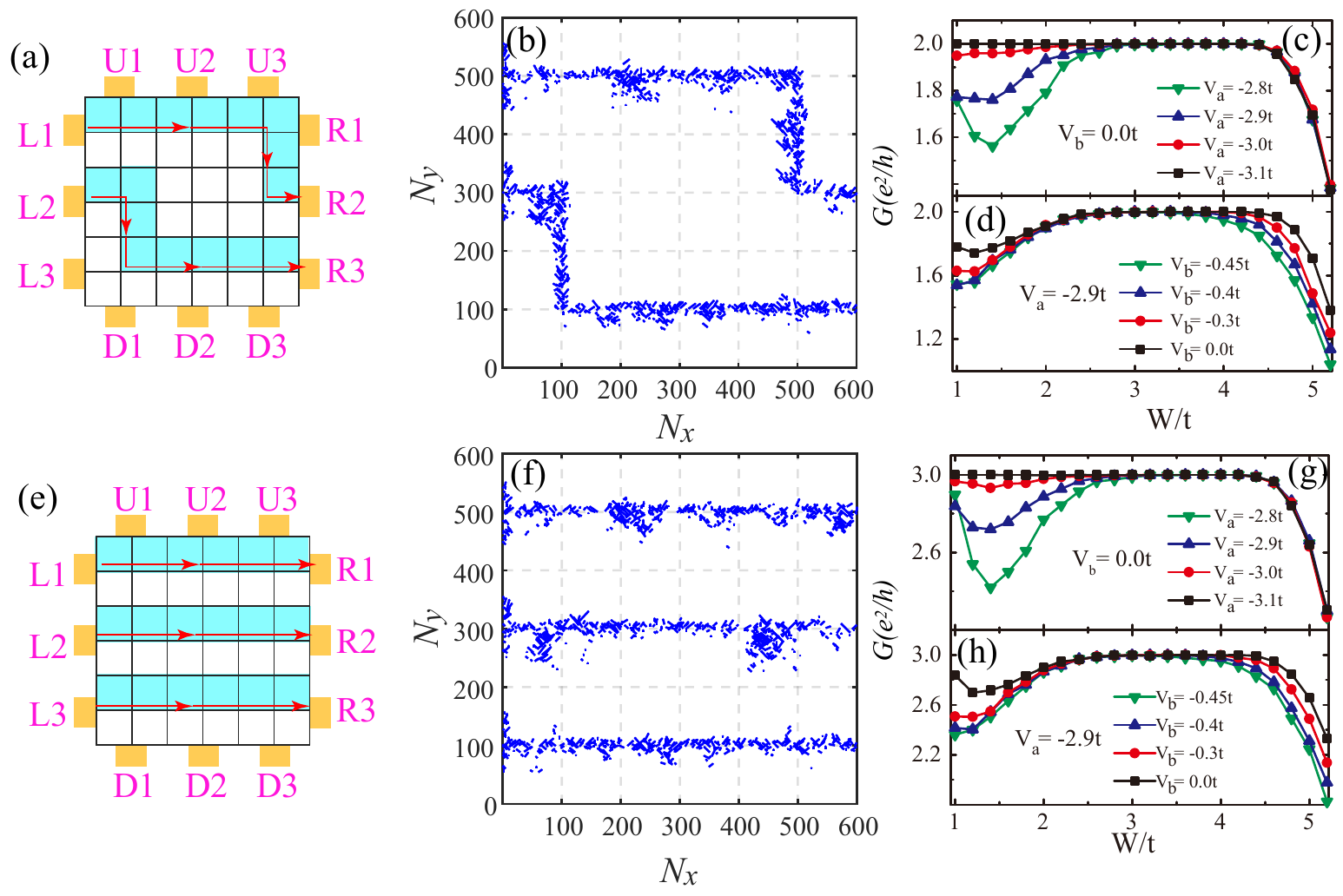}
	\caption{(Color online). (a) and (e) Schematic of programable circuit devices, which have two and three chiral interface channels, respectively. (b) and (f) Local current density distribution corresponding to cases (a) and (e) with gate voltage $ V_{a}=-2.8t $, $ V_{b}=0 $ and disorder strength $ W=3.5t $. (c)$ - $(d) and (g)$ - $(h) Conductances $ G $ $ vs $ $ W $ for different $ V_{a} $ and $ V_{b} $, respectively.}
	\label{f3}
\end{figure}

Three cases are investigated to illuminate the highly programable wires based on the dissipationless chiral-interface states. For simplicity, the standard gate voltage for the white (blue) blocks are chosen as $ V_{a}=-2.8t $ ($ V_{b}=0 $, if exists) with the Chern number $ C=0 $ ($ C=1 $) \cite{24}. In the first case, all gate voltages in blocks are set as $ V_{a} $ [see Fig. \ref{f2}(b)].
The entire sample belongs to the Anderson insulator with $C=0$. One can see from the typical local current density distribution (Fig. \ref{f2}(b)) and the conductance $G=0$ (Figs. \ref{f2}(c)), all logic devices in the circuit board are completely disconnected due to Anderson localization. In the second and the third cases, the gate voltages of blocks are programmed as configurations in Figs. \ref{f2}(d) and (g). The Chern number is $C=1$ in the blue blocks with $ V_{b}=0 $, and their difference between blue and white blocks guarantees the emergence of chiral interface channels labeled by the solid red lines with arrows. As shown in Fig. \ref{f2}(d) and (g), channel connecting device $ L3 $ ($ L2 $) and $ R2 $ ($ D2 $) is obtained. Therefore, it is appropriate to summarize that any two devices of a circuit can be switched into `on' and `off' by properly arranging the gate voltage $ V_{a}, V_{b} $.

In realistic samples, the disorder strength $ W $ and gate voltage $ V_{a}/V_{b} $ may deviate from the standard values because of the immature fabricating processes. To examine the robustness of wires in the programable circuit boards, we investigate the differential conductance $ G $ versus $ W $ for different $ V_{a}/V_{b} $.
As shown in Fig. \ref{f2}(c), the localization behaviors for states in Fig. \ref{f2}(b) are insensitive to the variation of $V_a$ and $W$. The plots in Fig. \ref{f2}(e)-(f) and (h)-(i) correspond to the configurations in Fig. \ref{f2}(d) and (g), respectively. The quantized $ G=e^{2}/h $ plateau exists for different gate voltages and a wide range of disorder strength, indicating the transport of these interface channels are robust against the variation of gate voltages and disorder strength. To be specific, when $ V_{a} $ ($ V_{b} $) takes the standard value $ -2.8t $ ($ 0t $), $ G=e^{2}/h $ plateau holds within a wide disorder strength range $ W\in [2.4t,\,4.4t] $. When the gate voltages on blue and white blocks deviate from the standard value, the plateau width shrinks slightly. Nevertheless, it is worth to note that as long as $ V_{a}\in [-3t,\,-2.6t] $ and $ V_{b}\in [-0.4t,\,0] $, one can always achieve the quantized $ G=e^{2}/h $ plateau with disorder strength $ W\in [2.4t,\,3.6t] $. This phenomenon indicates that the dissipationless programable circuit `wires' has a high fault tolerance feature for both gate voltage and disorder.

\begin{figure}[t]
	\centering
	\includegraphics[width=0.8\textwidth]{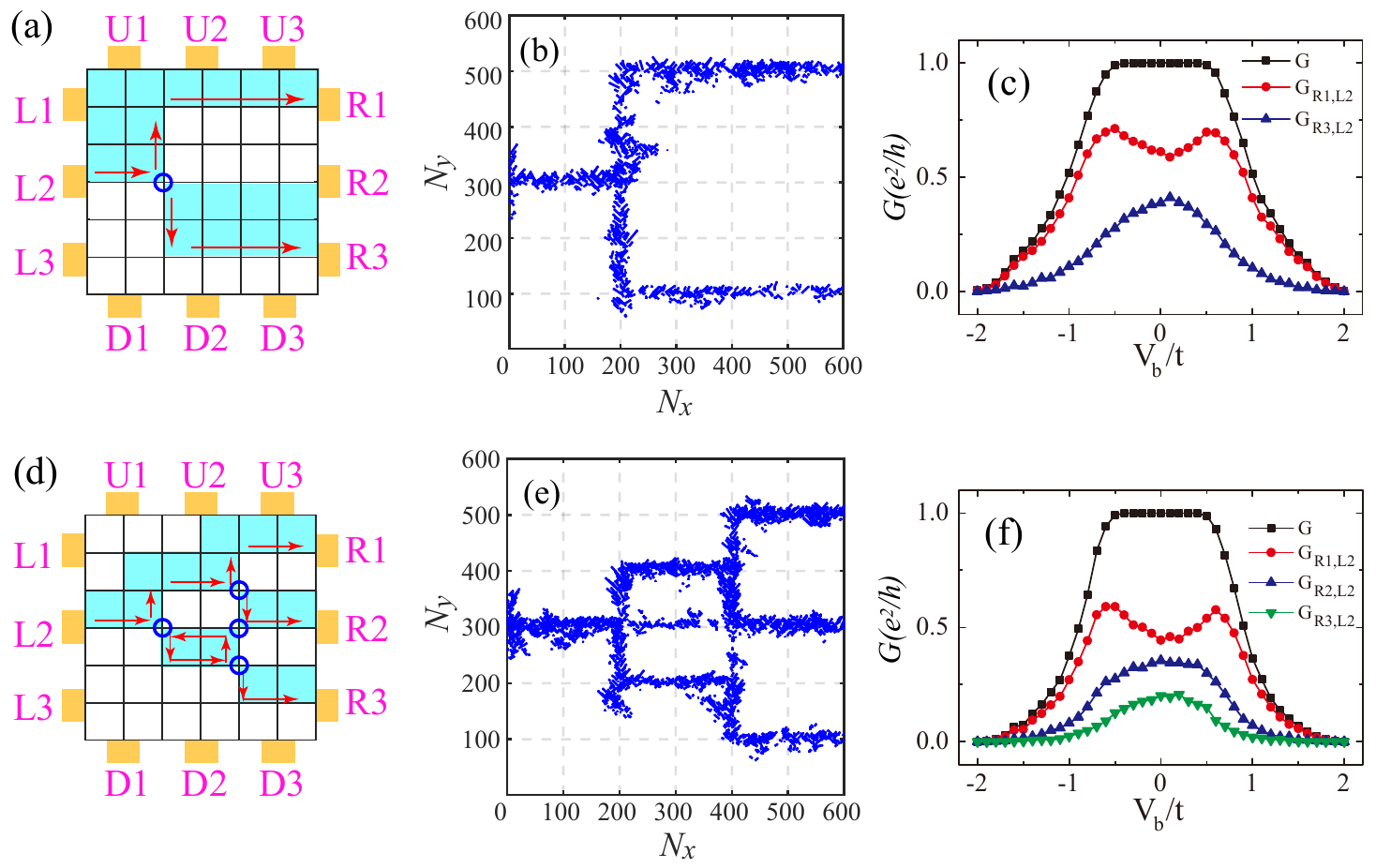}
	\caption{(Color online). (a)(d) The schematic diagram of gate voltage configuration to realize 2 and 3 branches of current in intergrated circuits devices, respectively. The blue circles denote bifurcation points of current. (b) and (e) Local current density distribution corresponding to (a) and (d), respectively, with $ V_{a}=0 $, $ V_{b}=-2.8t $ and disorder strength $ W=3.5t $. (c) and (f) The branch conductance $ G_{R1,L2} $, $ G_{R2,L2} $, $ G_{R3,L2} $ and the total conductance $ G $ $ vs $ $ V_{b} $ with $ V_{a}=0 $ and $ W=3.5t $.}
	\label{f4}
\end{figure}

Furthermore, we also study the gate voltage configurations shown in Fig. \ref{f3}(a) and (e). These two gate voltage configurations enable the simultaneous connection of two and three sets of devices [i.e., (L1-R2/L2-R3) in Fig. \ref{f3}(b) and (L1-R1/L2-R2/L3-R3) in (f)]. Figures \ref{f3}(c)-(d) and (g)-(h) manifest that the quantized $ G $ capture the double ($ G=2e^{2}/h $) and triple ($ G=3e^{2}/h $)  disspationless channels, respectively. Comparing to those in Fig. \ref{f2}, the parameter regions of $ W $ and $ V_{a}/V_{b}$ for quantized plateaus $ 2e^{2}/h $ ($ 3e^{2}/h $) are nearly unchanged. It means that parameters optimization is not needed when the single-wire is replaced by the multiple-wire.
Hence, such disordered-CIs-based programable wires possess fantastic potential applications due to their flexibility for device connection ($V$-tunable) and the robustness against parameter variation.

One can also program the chiral interface channels as current splitters, leading to a one-to-many connection between devices. Such a connection requires a more careful arrangement of gate voltages. For example, as shown in Fig. \ref{f4}(a) and (d), one-to-two and one-to-three connections are obtained. The corresponding local current density distributions [see Fig. \ref{f4}(b) and (e)] agree with the current flowing in the directions marked by red arrows.

To quantitatively study one-to-many wire connections, we pay more attention to the dependence of the current partition on the gate voltage $ V_{b} $ [see Fig. \ref{f4}(c) and (f)]. The differential conductance between different devices are studied, where $ G_{Ri,L2} $ signals the currents partitioned into device $ Ri $ from device $ L2 $. And the summation $G=\sum_{i\in{[1,2,3]}}G_{Ri,L2}$ gives the total conductance. From Fig. \ref{f4}(c) and (f), one find that $G$ still holds the quantized value $e^{2}/h$ with $ G_{Ri,L2} $ unquantized for $ V_{b}\in [-0.4t,\,0.4t] $. Specially, when $ V_{b} $ takes the standard gate voltage ($ V_{b}=0t $), all $ G_{Ri,L2} $ take the large values. It means that each branch wire is well connected. Furthermore, the conductance $ G_{Ri,L2} $ varies with $ V_{b} $, which provides a possible way to manipulate the current partition relations between branches. It's noteworthy that the quantized total conductance $G=e^2/h$ still manifests the dissipationless features of circuit boards in this case. According to all configurations studied in this section, one concludes that the external gates can be adopted to program the wires with arbitrary ($ n- $to$ -m $) connections in disordered CIs.

\begin{figure}[t]
	\centering
	\includegraphics[width=1\textwidth]{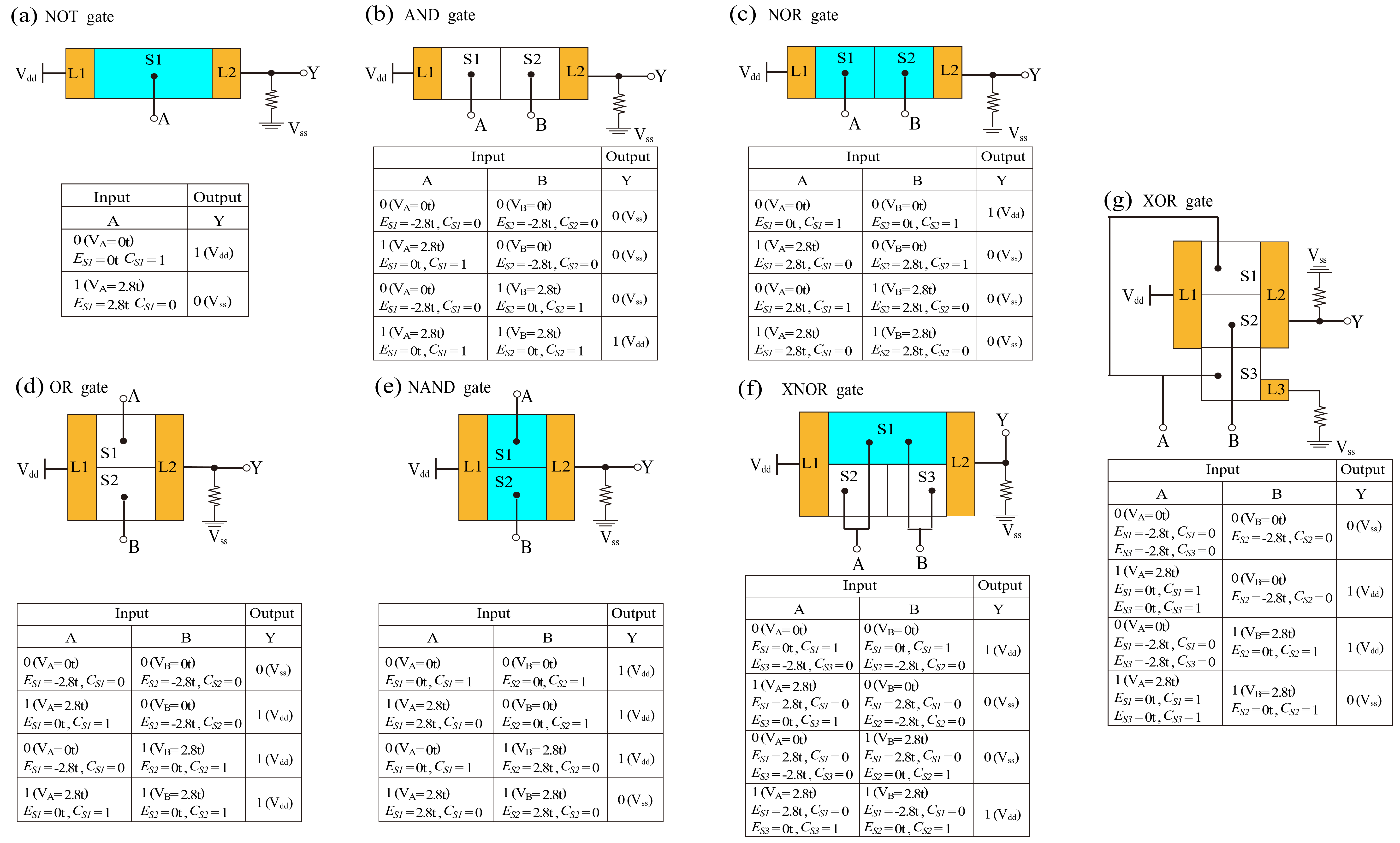}
	\caption{(Color online). (a)-(g) The schematic diagram of seven basic logic gates and the corresponding truth tables. The central region is divided into one, two or three blocks labelled by $ S1 $, $ S2 $ or $ S3 $, respectively. $ A $ and $ B $ denote voltage input terminals, connected to the gate on the corresponding block $ S1 $, $ S2 $ or $ S3 $. $ Y $ denotes voltage output terminal. $ V_{dd} $ and $ V_{ss} $ represent supply and source voltage. $ L1 $ and $ L2 $ or $ L3 $ represent the left and right terminals connected to the central region. When the voltage signal $ V_{A}=V_{B}=0 $, the initial Fermi energy in the blue block is $ E_{f}=0 $ (i.e., the CI with Chern number $ C=1 $), and in the white block is $ E_{f}=-2.8t $ (i.e., the Anderson insulator with $ C=0 $).}
	\label{f5}
\end{figure}

Apart from being used as programable dissipationless wire, disordered CIs also have a promising application in designing logic devices.
Generally, the logic devices are constructed by seven basic logic gates.
At present, the conventional logic gates have relatively complicated structures.
Taking the CMOS-based logic gate AND gate as an example, it is constructed by the combination of NAND gate and NOT gate. The NAND gate requires two pairs of PMOS and NMOS , while the NOT gate requires one pair of PMOS and NMOS. Comparing with the conventional logic gates, the structures of disordered-CIs-based logic gates are greatly simplified by taking advantage of gate-voltage-programable chiral edge and interface states.

Figures \ref{f5}(a)-(g) illustrate the construction schemes of seven basic logic gates and the corresponding truth tables.
Each of these seven logic gates has a central region, divided into several blocks. $ A $ and $ B $ denote two input terminals with voltage signals $ V_{A} $ and $ V_{B} $. $ V_{dd} $ and $ V_{ss} $ are supply and source voltages, respectively.
The initial Fermi energy in the blue block is $ E_{f}=0 $ (i.e., the Chern number $ C=1 $) and in the white block is $ E_{f}=-2.8t $ (i.e., the Anderson insulator with $ C=0 $).
For clarity, the Fermi energy $ E_{S1} $ ($ E_{S2} $ and $ E_{S3} $) and Chern number $ C_{S1} $ ($C_{S2} $ and $ C_{S3} $) for block $ S1 $ ($ S2 $ and $ S3 $) are shown in the truth table. The corresponding logic gate for different cases of voltage input are $ V_{A},V_{B} $. Importantly, we set $ V_{A/B}=V_{dd}=2.8t $ as the rated high level (i.e., logical `1') and $ V_{A/B}=V_{ss}=0t $ as the rated low level (i.e., logical `0').

Here, we take two of these seven logic gates as examples for a detailed analysis. Figure \ref{f5}(d) illustrate how the OR gate works. When $ V_{A}=V_{B}=0t $, $ S1 $ and $ S2 $ have the Fermi energy $ E_{S1}=E_{S2}=-2.8t $, and the corresponding Chern number is $ C_{S1}=C_{S2}=0 $. Thus, there is no current flowing into $ L2 $, and one has $ V_{Y}=V_{ss} $ (i.e., the logical operation: $ 0+0=0 $). When $ V_{A}=2.8t $,\,$ V_{B}=0t $, the Chern number of $ S1 $ becomes $ C_{S1}=1 $ ($ C_{S2} $ remains 0). The chiral interface channel emerges between $ S1 $ and $ S2 $. The current input from $ L1 $ will flow into $ L2 $ along this channel, and thus $ V_{Y}=V_{dd} $ (i.e., $ 1+0=1 $). For the last two cases $ V_{A}=0t $,\,$ V_{B}=2.8t $ (i.e., $ C_{S1}=0 $, $ C_{S2}=1 $) or $ V_{A}=V_{B}=2.8t $ (i.e., $ C_{S1}= C_{S2}=1 $), the current will flow into $ L2 $ along the lower edge of $ S1 $, and thus $ V_{Y}=V_{dd} $ (i.e., $ 0+1=1 $ and $ 1+1=1 $ ).

Similarly, Figure \ref{f5}(g) illustrates how the XOR gate works.
The central region is divided into three blocks $ S1 $, $ S2 $ and $ S3 $. Signal input terminal $ A $ is connected to the gates on both $ S1 $ and $ S3 $, $ B $ is connected to the gate on $ S2 $. In the case of $ V_{A}=V_{B}=0t $, the whole central region is an Anderson insulator ($ C_{S1}=C_{S2}=C_{S3}=0 $) with no current flowing through it (i.e., $ \bar{0}\cdot0 + 0\cdot\bar{0} =0$). When $ V_{A}=V_{B}=2.8t $, the Chern number of the whole central region is $ 1 $ ($ C_{S1}=C_{S2}=C_{S3}=1 $), the current will flow into $ L3 $ along the lower edge of region $ S3 $, rather than $ L2 $. Thus, $ L2 $ is always switched off for both cases with $ V_{Y}=V_{ss} $ (i.e., $ \bar{1}\cdot1 + 1\cdot\bar{1} =0$). However, when $ V_{A}=2.8t $,\,$ V_{B}=0t $ ($ V_{A}=0t $,\,$ V_{B}=2.8t $), the Chern number for $ S2 $ always differs by $ 1 $ compared to those for $ S1 $ and $ S3 $. The chiral interface channel emerges between $ S1 $ and $ S2 $ ($ S2 $ and $ S3 $). Generally, the current will flow into $ L2 $ along the interface channel between $ S1 $ and $ S2 $ or between $ S2 $ and $ S3 $ with $ V_{Y}=V_{dd} $ (i.e., $ \bar{1}\cdot0 + 1\cdot\bar{0} =1$ or $ \bar{0}\cdot1 + 0\cdot\bar{1} =1$). The rest of other logic gates can be analyzed in a similar manner.

Here, the feasibility of our logic gates proposal also originates from the spatial programable of chiral interface states. Compared to the conventional logic gates, the dissipationless transport features of chiral interface states can, in principle, significantly lower the power of logic gates in our cases. Specifically, we also analyze the logical operation of OR gate by investigating the conductance $ vs. $ disorder strength in the supplementary. The numerical results show that the logical operation is dissipationless and immune to disorder. Finally, the wires are compatible with the logic gates because they can be built based on one disordered CI sample. It will greatly simplify the integrated circuits fabricating process.

\section{Conclusion and Discussion}

In summary, we studied the realization of a programable integrated circuit based on disordered CIs. Due to the localization features, we propose the chiral interface channels as ideal wires. Significantly, dissipationless wires, which connect the arbitrary required devices, can be obtained by programming samples' gate voltage arrays. In addition to the wires, the disordered CIs can be utilized to construct the basic logic gates. Compared to the conventional counterparts, the simplified logic gates' structures will greatly promote the compaction of integrated circuits. Moreover, combined with the dissipationless characteristics of the wires as well as the logic devices, our proposal will enable circuits with lower power consumption, higher integration, and reliability.

Notably, since the Anderson phase transition is basic and only determined by the system's dimension and symmetry ensemble, our results apply to all 2D CIs that exhibit a direct transition from the CI to the normal insulator. Furthermore, to construct such a programable integrated circuit, one only needs two discrete standard gate voltages ($ V_{a} $ and $ V_{b} $) corresponding to the topologically nontrivial insulator and the normal insulator, respectively. The proposal is still available if $ V $ can drive the transition from CI to a band insulator in some systems\cite{38,39}. This also means that it does not matter if there exists a metallic phase during the Anderson transition, since one can skip the voltage windows corresponding to the metallic phase \cite{40}. Therefore the CIs can be further broadened to other topological nontrivial systems such as the QSHE, etc. \cite{41,42,43}. Finally, we discuss the scaling of the integrated circuits, which is limited by the size of gate blocks. For a typical QAH system with Fermi velocity $ \nu_{F}\approx 5\times 10^{5} m/s $ and energy gap $ \Delta \approx 20meV$, the decay length is $ \lambda \approx \frac{\hbar \nu_{F}}{\Delta} \sim 50nm $. The limited scaling can be smaller than $ 0.5 \mu m $, since the gate block will work well if the size is one order of magnitude larger than $ \lambda $.
Experimentally, our proposal only depends on the size scale for the Anderson transition in CIs.
Fortunately, the Anderson localization has been recently observed in $ MnBi_{2}Te_{4} $ samples of $ 20\mu m \times 20\mu m $ \cite{15,44}, indicating the proposed integrated circuits are feasible under the state-of-the-art experimental techniques.

\section{Method}
The non-equilibrium Green's function method \cite{36,37} is adopted to simulate the transport properties with  $G_{pq}=\frac{e^{2}}{h}Tr[\Gamma_{p}G^r\Gamma_{q}G^a] $ the conductance between terminals $p$ and $q$.
$ G^{r/a}=[E_F\pm i0^+-H- \sum_{p}\Sigma_p]^{-1} $ are the retarded$ /$advanced Green's function, and $ \Gamma_{p}=i[\Sigma_p-\Sigma_p^\dagger] $ is a linewidth function with $\Sigma_p$ the self-energy of terminal $p$.
The local current flow vector is calculated by:
\begin{equation}
	\textbf{J}_{i\rightarrow j}=\frac{2e^2V}{h}Im[H_{i,j}(G^r\Gamma_LG^a)_{j,i}].
\end{equation}
$H_{i,j}$ is the coupling matrix between $i$ and $j$ sites. The local current flow vector for site $i$ is $\textbf{J}_i=[\textbf{J}_{i\rightarrow i+\hat{x}}+\textbf{J}_{i\rightarrow i+\hat{y}}]$.

\section{Additional Information}
The Supplementary Information for this paper is available at http://xxxxx.
	
	Logical operation for OR gate.
	
\section{Author Contributions}
J. H initialed the ideal from the discussion with Z. B. W and integrated the research. B.L.W and Z. Q. Z performed the calculations. B. L. W, Z. Q. Z and H. J wrote the manuscript. All authors have given approval to the final version of the manuscript.

\section{Notes}
The authors declare no competing financial interest.

\section{Acknowledgement}

We are grateful to Chui-Zhen Chen, Rui-Chun Xiao, and  Jinsong Zhang for helpful discussion. This work was supported by National Basic Research Program of China (Grant No. 2019YFA0308403), NSFC under Grant No. 11822407, and a Project Funded by the Priority Academic Program Development of Jiangsu Higher Education Institutions.

\section{Supplementary Material: LOGICAL OPERATION OF $OR$ $GATE$}



\setcounter{equation}{0}
\setcounter{figure}{0}
\setcounter{table}{0}
\setcounter{page}{1}
\makeatletter
\renewcommand{\theequation}{S\arabic{equation}}
\renewcommand{\thefigure}{S\arabic{figure}}
\renewcommand{\bibnumfmt}[1]{[S#1]}
\renewcommand{\citenumfont}[1]{S#1}

We take \emph{OR gate} as an example and calculate the conductance $ G $ versus disorder strength $ W $ for four cases of voltage input ($ V_{A} $,\,$ V_{B} $), as shown in Fig. S1.
When the output level is logical `1' [i.e., (i) $ V_{A}=2.8t $,\,$ V_{B}=0t $; (ii) $ V_{A}=0t $,\,$ V_{B}=2.8t $; (iii) $ V_{A}=2.8t $,\,$ V_{B}=2.8t $], the quantized $ G=e^{2}/h $ plateau exists within a wide range of disorder strength $ W\in [2.4t,4.4t] $.
When the output level is logical `0' (i.e., $ V_{A}=0t $,\,$ V_{B}=0t $), $ G=0 $ for the same disorder region.
It means that the logical operations of \emph{OR gate} are dissipationless and robust against disorder.

\begin{figure}[h]
	
	\centering
	\includegraphics[width=0.5\textwidth]{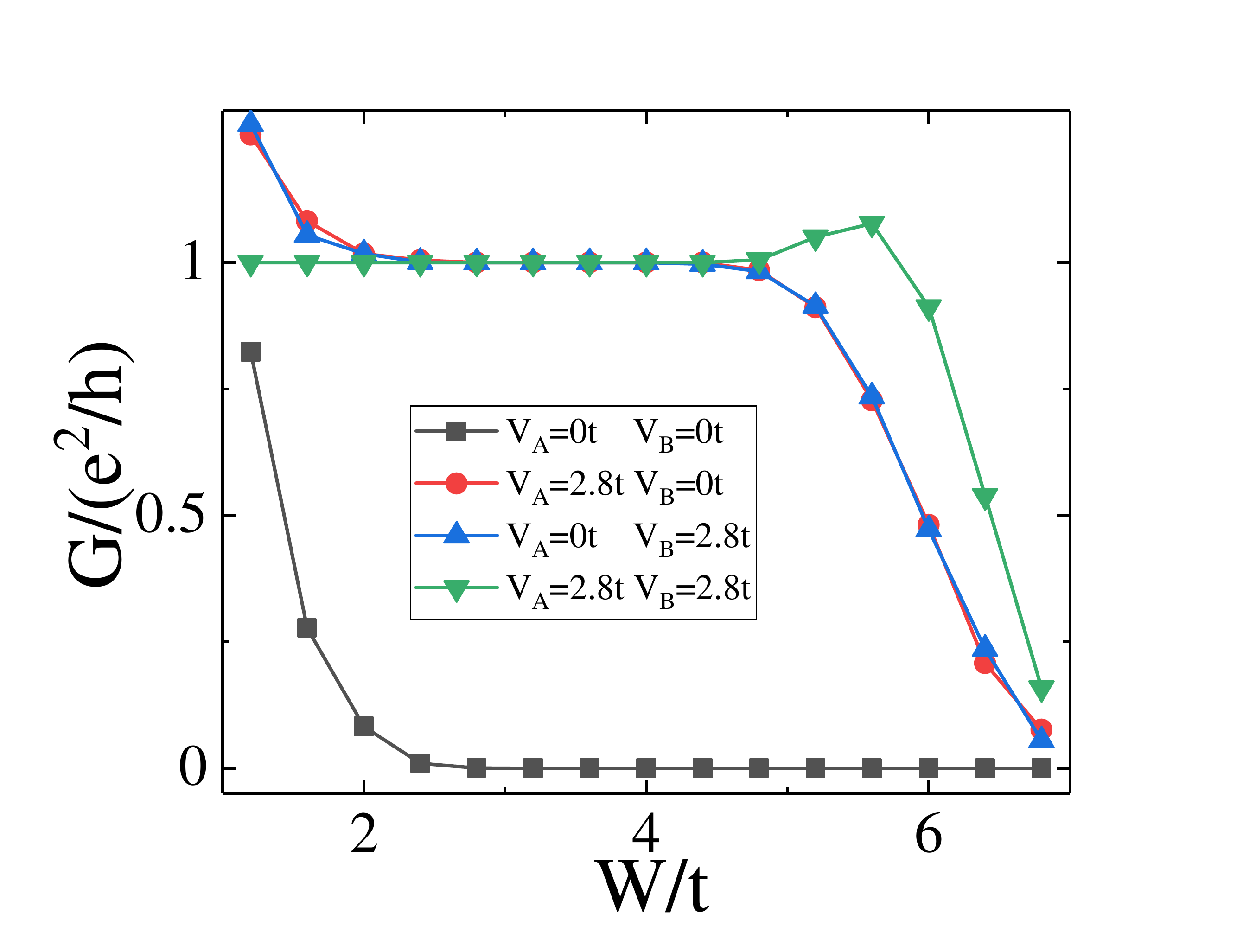}
	\caption{(Color online). Differential conductances $ G $ versus disorder strength $ W $ for \emph{OR gate}. }
	\label{S1}
\end{figure}

\end{document}